\pdfoutput=1
\documentclass[aps,prd,twocolumn,superscriptaddress,preprintnumbers,floatfix,nofootinbib]{revtex4-2}

\usepackage{graphicx}
\usepackage{amsmath}
\usepackage[caption=false]{subfig}
\usepackage{siunitx}
\usepackage{placeins}
\usepackage{color}
\usepackage{standalone}
\usepackage{dcolumn}
\usepackage{tensor}
\usepackage{bm}
\usepackage{microtype}
\usepackage{etoolbox}
\usepackage{amssymb}
\usepackage{mathrsfs}
\usepackage{accents}
\usepackage[normalem]{ulem}
\usepackage[dvipsnames]{xcolor}
\usepackage[colorlinks,urlcolor=NavyBlue,citecolor=NavyBlue,linkcolor=NavyBlue,pdfusetitle]{hyperref}
\usepackage[all]{hypcap}
\usepackage[inline]{enumitem}
\usepackage[utf8]{inputenc}
\usepackage{lipsum}
\usepackage{booktabs}
\usepackage{wasysym}

\usepackage{array}

\newcommand{\beq}{\begin{equation}}
\newcommand{\eeq}{\end{equation}}

\newtoggle{commentsoff}
\togglefalse{commentsoff}

\ifdefined\nocomments
    \toggletrue{commentsoff}
\fi

\definecolor{dodgerblue}{HTML}{1E90FF}

\iftoggle{commentsoff}{
  \usepackage[final,commandnameprefix=ifneeded]{changes}

  \newcommand*{\mi}[1]{}
  \newcommand*{\wf}[1]{}
  \newcommand*{\kc}[1]{}
  \newcommand{\sv}[1]{}
  \newcommand{\ep}[1]{}
  
  \newcommand*{\todo}[1]{}
  \newcommand*{\warn}[1]{}

}{
  \usepackage[commentmarkup=uwave,commandnameprefix=ifneeded]{changes}

  \newcommand*{\mi}[1]{{\color{magenta} [{\bf MAX}: #1]}}
  
  \newcommand*{\kc}[1]{\textcolor{ForestGreen}{[\textbf{KC}: #1]}}
  \newcommand{\sv}[1]{\textcolor{BurntOrange}{sv: #1}}
  \newcommand{\ep}[1]{\textcolor{RoyalPurple}{[\textbf{EP}: #1]}}
  
  \newcommand*{\warn}[1]{{\color{red} [{\bf WARNING}: #1]}}
  \newcommand*{\todo}[1]{{\color{red} [{\bf TODO}: #1]}}

  \newcommand*{\wf}[1]{\textcolor{violet}{[\textbf{WF:} #1]}}
}

\definechangesauthor[name={Max}, color=magenta]{MI}
\definechangesauthor[name={Ethan}, color=RoyalPurple]{EP}

\graphicspath{{./fig/}}

\newcommand{\dcc}{LIGO-P}

\begin{document}


\title{The impact of selection biases on tests of general relativity with gravitational-wave inspirals}

\newcommand{\CCA}{\affiliation{Center for Computational Astrophysics, Flatiron Institute, 162 5th Ave, New York, NY 10010}}
\newcommand{\CIT}{\affiliation{Department of Physics, California Institute of Technology, Pasadena, California 91125, USA}}
\newcommand{\CITLab}{\affiliation{LIGO Laboratory, California Institute of Technology, Pasadena, California 91125, USA}}
\newcommand{\MITLab}{\affiliation{LIGO Laboratory, Massachusetts Institute of Technology, 185 Albany St, Cambridge, MA 02139, USA}}
\newcommand{\MIT}{\affiliation{Department of Physics and Kavli Institute for Astrophysics and Space Research, Massachusetts Institute of Technology, 77 Massachusetts Ave, Cambridge, MA 02139, USA}}
\newcommand{\Birmingham}{\affiliation{School of Physics and Astronomy and Institute for Gravitational Wave Astronomy, University of Birmingham, Edgbaston, Birmingham, B15 2TT, United Kingdom}}

\newcommand{\StonyBrook}{\affiliation{Department of Physics and Astronomy, Stony Brook University, Stony Brook NY 11794, USA}}

\author{Ryan Magee}
\email{rmmagee@caltech.edu}
\CIT
\CITLab

\author{Maximiliano Isi}
\email{misi@flatironinstitute.org}
\CCA

\author{Ethan Payne}
\email{epayne@caltech.edu}
\CIT
\CITLab

\author{Katerina Chatziioannou}
\email{kchatziioannou@caltech.edu}
\CIT
\CITLab

\author{Will M.~Farr}
\email{wfarr@flatironinstitute.org}
\CCA
\StonyBrook

\author{Geraint Pratten}
\email{g.pratten@bham.ac.uk}
\Birmingham

\author{Salvatore Vitale}
\email{salvo@mit.edu}
\MITLab
\MIT

\hypersetup{pdfauthor={Magee, Isi, Payne, Chatziioannou, Pratten, Vitale, Farr}}

\date{\today}

\begin{abstract}
Tests of general relativity with gravitational wave observations from merging compact binaries continue to confirm Einstein's theory of gravity with increasing precision.
However, these tests have so far only been applied to signals that were first confidently detected by matched-filter searches assuming general relativity templates.
This raises the question of selection biases: what is the largest deviation from general relativity that current searches can detect, and are current constraints on such deviations necessarily narrow because they are based on signals that were detected by templated searches in the first place?
In this paper, we estimate the impact of selection effects for tests of the inspiral phase evolution of compact binary signals with a simplified version of the \textsc{GstLAL} search pipeline.
We find that selection biases affect the search for very large values of
the deviation parameters, much larger than the constraints implied by the
detected signals.  Therefore, combined population constraints from
confidently detected events are mostly unaffected by selection biases,
with the largest effect being a broadening at the $\sim10$\%
level for the $-1$PN term.  These findings suggest that current
population constraints on the inspiral phase are robust without factoring
in selection biases. Our study does not rule out a disjoint, undetectable
binary population with large deviations from general
relativity, or stronger selection effects in other tests or search
procedures.
\end{abstract}


\maketitle



\section{Introduction}

Gravitational wave (GW) signals detected by LIGO~\cite{TheLIGOScientific:2014jea} and Virgo~\cite{TheVirgo:2014hva} have provided otherwise-inaccessible constraints on deviations from general relativity (GR) in the dynamical and strong-field regimes~\cite{TheLIGOScientific:2016src,Yunes:2016jcc, LIGOScientific:2019fpa, LIGOScientific:2020tif, LIGOScientific:2021sio}.
When considered in aggregate, the set of detected binary black hole (BBH)
signals is fully consistent with the null hypothesis of quasicircular
mergers in vacuum GR. However, existing constraints apply only to
signals that have been confidently detected and identified as compact
binaries by pipelines based on GR. Even though generic searches
exist~\cite{Klimenko:2015ypf,Cornish:2020dwh,LIGOScientific:2016fbo,LIGOScientific:2016kum,KAGRA:2021tnv},
all current BBH signals have been detected with search pipelines that are
based on templates produced within Einstein's theory. It remains possible that there
exist binaries whose signals depart from GR but have been selected against
by searches \cite{Chia:2020psj,Chia:2023tle,Narola:2022aob}. This raises two interrelated questions: (i) what is the largest
deviation from GR that current searches can detect? and (ii)
are current constraints on deviations from GR artificially narrow because
they are based on signals that were detected in the first place?

Answering these questions amounts to quantifying the \emph{selection biases}
that modulate the probability of signal detection as a function of its
parameters.  The impact of regular binary parameters within GR---such as black
hole (BH) masses or spins---can be approximated through their influence on the
expected signal-to-noise ratio (SNR) of a given
signal~\cite{LIGOScientific:2018jsj,LIGOScientific:2020kqk}, or more robustly
by assessing the performance of the search pipeline on simulated
signals~\cite{KAGRA:2021duu}. The resulting selection function is an
indispensable ingredient in inferring the astrophysical distributions of the
detected
events~\cite{LIGOScientific:2018jsj,LIGOScientific:2020kqk,KAGRA:2021duu}.
While this effect is well understood for GR parameters, the selection on
beyond-GR parameters is currently largely unknown and generally unquantified.
Nevertheless, studies under specific models suggest searches have nonnegligible selection for
sufficiently large deviations~\cite{Chia:2020psj,Chia:2023tle,Narola:2022aob}. 

In the absence of a quantified selection function for tests of GR, current
constraints are restricted to assessing agreement of the population properties of detected events with GR. 
Such an analysis can be performed without reference to any
specific alternative theory of gravity by inferring the general shape of the
population of deviations using hierarchical
inference~\cite{James:1961,Lindley:1972,Efron:1977,Rubin:1981}. This procedure
can detect anomalies in a collection of signals even if the deviation manifests
differently for each individual event~\cite{Zimmerman:2019wzo,Isi:2019asy,Isi:2022cii}. However, without selection effects, this procedure does not infer the \emph{intrinsic} population of deviations, which could contain undetectable
signals~\cite{Isi:2019asy,LIGOScientific:2020tif,LIGOScientific:2021sio}.
Furthermore, if selection biases are strong, these population constraints do not formally correspond to the \emph{detected population} either on account of detector noise~\cite{Essick:2023upv}.
This concern also extends to cases in which events can be combined by simply multiplying likelihoods for a shared deviation parameter.

In this paper, we study the selection function within 
template-based search pipelines for parameterized tests of the inspiral phasing
parameters~\cite{Yunes:2009ke,Li:2012a,Li:2012b,Agathos:2013upa}.
Among the wide array of possible GR tests, we focus on post-Newtonian (PN)
modifications to the waveform phasing, $\varphi(f)$, due to anomalous
dynamics~\cite{Cornish:2011ys,Li:2011cg,Li:2011vx,Agathos:2013upa,Sampson:2013jpa,Sampson:2013lpa,Meidam:2014jpa,DelPozzo:2011pg,Ghosh:2016qgn,TheLIGOScientific:2016pea,Meidam:2017dgf,Ghosh:2017gfp,Brito:2018rfr},
which could arise from corrections to the theory or due to exotic sources
following other nonstandard physics, such as BH mimickers. We use the
deviation parameters $\delta \varphi_{i}$, where $i/2$ denotes the associated
 PN order.
 We focus on PN modifications as they are one of the flagship tests of GR with LIGO, Virgo, and KAGRA \cite{KAGRA:2020tym}, and their effect is to modify the full inspiral, which dominates the detectability of all but the most massive systems.
The latter can more easily be detected by theory-agnostic burst pipelines,
potentially reducing the expected impact of selection biases induced by
deviations from GR.

We generate simulated signals (also called \emph{injections}) and recover them
with a simplified version of the \textsc{GstLAL}
pipeline~\cite{Messick:2016aqy,Sachdev:2019vvd,2021SoftX..1400680C,Tsukada:2023edh}in Sec.~\ref{sec:methods}.  Rather than evaluating the computationally
expensive likelihood ratio that would normally be computed by \textsc{GstLAL}
as a detection statistic, we approximate detection efficiency with a proxy
ranking statistic based on the recovered SNR and an autocorrelation-based
consistency check.  In Sec.~\ref{sec:singleinj} we find that, under these
circumstances, selection biases affect the detectability of signals only for
very large values of the deviation parameters. 
These values are significantly higher than the precision achieved by current tests; we therefore expect that incorporating selection effects in population
inference will have a minimal impact on the resulting constraints.

Armed with the results from our injection campaign, we confirm this expectation by enhancing existing hierarchical tests of GR \cite{LIGOScientific:2021sio} with a selection factor, and compute the resulting astrophysical distribution of deviation parameters in Sec.~\ref{sec:population}. 
We parametrize the deviation population with a Gaussian and infer its mean and standard deviation while taking into account selection effects.
Following~\cite{Payne:2023kwj}, we simultaneously model the astrophysical distribution of the binary component masses.
For most phase deviation terms we consider, the inferred astrophysical distributions for beyond-GR parameters are identical to those obtained by ignoring the GR selection effects.
We recover the strongest impact for the $-1$PN term, where incorporating selection effects widens the inferred population distribution by 10\%.
We therefore conclude that the quantitative impact of ignoring selection effects in tests of GR with GW inspirals is small.

This conclusion may be surprising given the crucial role of selection effects in estimating, for example, the mass
distribution of BBHs. 
The crucial difference between deviation parameters and BBH masses is that the former population is inferred to be intrinsically very narrow as all events are consistent with a vanishing deviation. 
Indeed, after a dozen high-significance BBHs, the population for all deviation parameters inferred from LIGO-Virgo data is already narrower than the impact of selection effects.
As more events are detected (and assuming they remain consistent with GR), the inferred deviation population will continue to narrow, making selection effects even less relevant.
In other words, selection effects do exist in the population, but their impact is only appreciable for deviation values that are already ruled out.
Other population distributions, such as those for the mass and spin, are not inherently narrow and selection effects remain important no matter how many events are detected.
These considerations suggest that our conclusions only apply under the assumption that all events come from a narrow, unimodal population of deviation parameters. They do not rule out a disjoint population with deviations large enough to remain hidden to searches; such extreme non-GR signals can only be ruled out with a dedicated search~\cite{Chia:2020psj,Chia:2023tle,Narola:2022aob}.
We further this argument in our concluding remarks, Sec.~\ref{sec:conclusions}.

\section{Estimating the matched-filter selection function for signals with GR deviations}
\label{sec:methods}

In this section, we describe the procedure for quantifying the effect of GR deviations on the GW selection function. 
In summary, we follow the standard practice of estimating detection efficiency by simulating a large set of signals (Sec.~\ref{sec:injections}), analyzing them with a detection pipeline (Sec.~\ref{sec:gstlal}), and determining which signals are detectable (Sec.~\ref{sec:efficiency}).

\subsection{Injection Set}
\label{sec:injections}

We start with the publicly available set of $156\thinspace878$ BBH injections
associated with GWTC-3, which target only GR parameters~\cite{o3-selection}; we
leave detailed explorations of binary neutron stars and neutron-star--black-hole binaries to future work. In this injection set, the primary and
secondary binary masses are distributed as $p(m_1) \propto m_1^{-2.35}$ and
$p(m_2|m_1) \propto m_2$ and bounded, in the source frame, such that $2\,
M_\odot < m_2 \leq m_1 < 100\, M_\odot$; the BH spins are isotropically
distributed with uniformly distributed magnitudes $|\chi_{1,2} | \leq 0.998$. Further specifics of
the within-GR population are described in Table XII
of~\cite{LIGOScientific:2021djp}. The simulations are generated using a baseline
\textsc{IMRPhenomPv2} waveform
approximant~\cite{Husa:2015iqa,Khan:2015jqa,Hannam:2013oca}, which includes the
effects of spins misaligned with the orbital angular momentum.
We implement deviations from GR using the \textsc{TIGER} framework~\cite{Li:2012a,Li:2012b,Agathos:2013upa}, as in~\cite{LIGOScientific:2021sio}.

To reduce the computational burden on the original GWTC-3
analysis~\cite{LIGOScientific:2021djp}, these injections have already been
selected against a minimum optimal network signal-to-noise ratio (SNR)
threshold of $ 6$. The network SNR was calculated by adding the
LIGO-Livingston and LIGO-Hanford SNRs in quadrature.  Systems with a lower
optimal network SNR are considered ``hopeless" for detection.  To
further enhance computational efficiency, we only consider BBHs that have
optimal LIGO Livingston SNRs $\geq 6$ and redshifted total masses below $300\,
M_\odot$.  For our purposes, restricting the total mass injected has negligible
effect due to the additional inspiral SNR selection criterion typically applied
in PN tests of GR~\cite{LIGOScientific:2021sio, Payne:2023kwj}; 
we return to this in Sec.~\ref{sec:population}.  These initial cuts result in $84\thinspace119$ injections.

To measure the selection bias against beyond-GR populations, we perturb the
inspiral phasing of the injections and recover them with an approximation of
the \textsc{GstLAL}-based inspiral pipeline described in Sec.~\ref{sec:gstlal}.
Following the standard parametrized post-Einsteinian test~\cite{Yunes:2009ke},
we perturb each PN order and repeat the analysis separately.  Each simulation
is assigned a random fractional\footnote{In GR, the coefficients corresponding to the -1PN and 0.5 PN terms are exactly zero. $\delta \varphi_{-2}$ and $\delta \varphi_{1}$ therefore represent absolute deviations.} deviation drawn from a uniform distribution
with bounds $\pm 0.1$, $\pm 1$, $\pm 5$, $\pm 3$, $\pm 2$, $\pm
15$, $\pm 5$, $\pm 10$, $\pm 50$, and $\pm 30$ for the $\delta \varphi_{-2}$, $\delta \varphi_{0}$, $\delta
\varphi_{1}$, $\delta \varphi_{2}$, $\delta \varphi_{3}$, $\delta
\varphi_{4}$, {$\delta \varphi_{5l}$, $\delta \varphi_{6}$, $\delta
\varphi_{6l}$, and $\delta \varphi_{7}$, respectively, where the ``$l$"
subscript denotes the logarithmic phase terms. The bounds are chosen such that the inferred deviations from individual events are entirely covered by the selection. We only vary one coefficient at
a time to match the analysis usually applied to actual data
\cite{LIGOScientific:2021sio}. 
This procedure results in one
BBH injection set per PN order, each containing the
same number of BBHs with identical GR parameters, differing only in the order and
strength of the random GR deviations. After specifying injection parameters, 
we generate a corresponding waveform using the \textsc{IMRPhenomPv2} approximant and add it to the data stream of a single detector.  We space the
simulated signals $7$ seconds apart through a single stretch of data collected
in the LIGO Livingston detector during April of
2019 with global-positioning-system (GPS) times in the range $[1239641219\, {\rm s}, 1240334066\,{\rm s}]$~\cite{LIGOScientific:2023vdi}.

\subsection{Detection criterion and efficiency}
\label{sec:gstlal}
\label{sec:efficiency}

\begin{figure}

\includegraphics[width=\columnwidth]{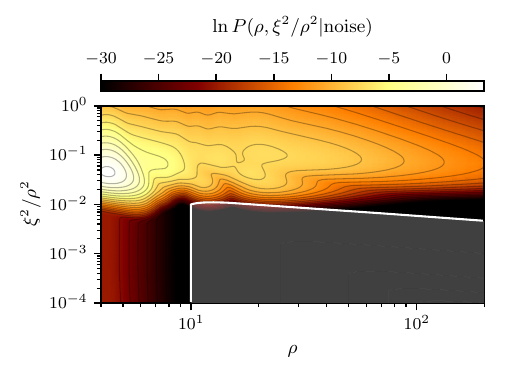}
\caption{\label{fig:search_background} A representative background distribution
for BBHs collected for the LIGO Livingston detector. The
background is parameterized in $\xi^2/\rho^2$ vs $\rho$ space. Regions
with high $\ln P$ indicate where noise is most likely (brighter color). The shaded contour enclosed by a white edge corresponds to our detection criterion, $\bar{\rho}\geq 10$. This region is
largely separate from the collected background.}
\end{figure}

We analyze the injection sets with a simplified infrastructure based on \textsc{GstLAL}, one of the matched-filter-based search pipelines presently used
to search for GWs from compact binaries~\cite{Sachdev:2019vvd,
Hanna:2019ezx, Messick:2016aqy,Tsukada:2023edh,Allen:2005fk, Allen:2004gu,
Canton:2014ena,
Usman:2015kfa,Nitz:2017svb,Adams:2015ulm,Aubin:2020goo,chu2017low,Venumadhav:2019tad}.
Matched-filter based search
pipelines discretely sample the GR-based signal manifold to create template banks of
possible signals. The discretization results in a $1\% {-} 3\%$ loss of SNR over the parameter space covered by the bank~\cite{2021PhRvD.103h4047M,Sakon:2022ibh}.
Pipelines presently restrict their searches to emission from sources with spin
angular momenta aligned with the orbital angular momenta, and
therefore neglect the impact of precession or higher-order angular modes; the
signal loss incurred for these systems is, therefore, larger. We specifically
consider the \textsc{GstLAL}-based matched filtering pipeline for its signal consistency check and because it most
densely sampled the signal space in LIGO-Virgo's third observing run (O3), and
thus had the minimum expected SNR loss from discreteness. For BBHs, the
\textsc{GstLAL} bank used an effective-one-body model of the GW emission,
\textsc{SEOBNRv4\_ROM}~\cite{Bohe:2016gbl}. The specific structure and maximum
SNR loss 
of \textsc{GstLAL}'s template bank is described in Table II of the
GWTC-2 publication~\citep{LIGOScientific:2020ibl}.

Pipelines correlate waveforms from the template bank with the
data collected in each detector to produce an SNR time series. 
Peaks in the SNR time series, called triggers, are checked for coincidence across detectors, and
are then ranked according to the pipeline's detection statistic. \textsc{GstLAL}'s ranking statistic is the likelihood
ratio $\mathcal{L}$, defined in~\citep{Cannon:2015gha,Tsukada:2023edh}, which relates the
probability of observing a set of parameters under the signal hypothesis to
that of the instrumental-noise hypothesis. This quantity is a function of a number of factors: the set of instruments
participating in a detection, the matched-filter SNR, a signal-based-veto parameter, the
event time and phase in the frame of each detector,
and the masses and spins of the identifying template.
In general, it is computationally expensive to accurately estimate the
background of the search and recover simulated signals via $\mathcal{L}$. Since
no background for O3 is publicly available, and to
minimize the analysis cost, we instead employ an approximate detection
statistic $\bar{\rho}$ that weights the measured SNR by a signal consistency check~\cite{LIGOScientific:2011jth,Babak:2012zx}, namely
\begin{equation} 
\bar{\rho} = \frac{\rho}{\left[\frac{1}{2}\left(1 + \max(1, \xi^2)^3\right)\right]^{1/5}}\, , 
\label{eq:detstatistic}
\end{equation}
where $\rho$ is the matched-filter SNR and $\xi^2$ is a signal consistency test defined from the autocorrelation as 
\begin{equation} 
\xi_j^2 = \frac{\int_{-\delta t}^{\delta t}\mathrm{d}t \, | z_j(t) - z_j(0)\, R_j(t) |^2 \,}{\int_{-\delta t}^{\delta t}\mathrm{d}t \, (2 - 2 \left| R_j(t) \right|^2) \,}\,,
\label{eq:consistency-def}
\end{equation}
where $z_j$ and $R_j$ denote the complex SNR and autocorrelation of template
$j$, respectively, and the integrand in the denominator is the expectation
value in Gaussian noise~\cite{Messick:2016aqy}. We compute a value of $\xi^2$
for each trigger by integrating Eq.~\eqref{eq:consistency-def} over a small
window of time $\pm \delta t$, centered about the trigger. We use $\delta t =
0.17\, {\rm s}$ ($\delta t = 0.34\, {\rm s}$) for templates with chirp masses greater (less)
than 15 $M_\odot$, which was also done in production by the full
\textsc{GstLAL} pipeline. When the observed strain data closely matches the
template $j$, then $\bar{\rho} = \rho$.  For each BBH injection, we compute the
matched-filter SNR $\rho$ and
$\xi^2$ value against the \textsc{GstLAL} template bank. Since signals
generally match with multiple templates in a bank, we perform the same
data reduction clustering as the \textsc{GstLAL} pipeline does in GWTC-3.
We discard triggers within $0.1$\,s of other triggers with a larger
$\bar{\rho}$ value, breaking ties by $\rho$.

Since we consider the response in only a single detector, we
conservatively set a detection threshold of $\bar{\rho} \geq 10$. This
choice is motivated by the fact that significant candidates from GWTC-2 and
GWTC-3 were identified for network SNR $\rho_\mathrm{net} \gtrsim 10$, which
typically corresponded to events with single detector SNRs $\rho_\mathrm{H}
\sim \rho_\mathrm{L} \sim 7$. As we only filter a single detector, we assert
that a signal in a single detector with $\rho=10$ will have approximately the
same significance as a signal observed in multiple detectors with
$\rho_\mathrm{net}=10$.  We further assert that our
proxy detection statistic threshold is approximately equivalent to the
false-alarm-rate (FAR) threshold of $\mathcal{O}(10^{-3} /
\mathrm{yr})$ adopted in past tests of GR~\cite{LIGOScientific:2019fpa,
LIGOScientific:2020tif, LIGOScientific:2021sio}. This choice is conservative
for our study in that a
weaker detection criterion could only reduce the \emph{detection} bias, i.e., it could
only increase the fraction of signals that are detected by the pipeline.

Although we use an abbreviated version of the detection pipeline, 
we argue that the resulting selection function is a good approximation for the full
selection effect for the following reasons:
\begin{enumerate}

\item The threshold of $\bar{\rho}>10$ selects triggers that are disjoint from the
background typically collected by the search. Triggers that meet this
criterion exist in the shaded contour shown in
Fig.~\ref{fig:search_background}, which is cleanly off a representative background
observed by the search. In other words, $\bar{\rho}>10$ implies vanishing support from the background.

\item
In addition to the background, $\mathcal{L}$ contains a signal term that
we do not explicitly take into account here. This is justified because, in
the $\bar{\rho}>10$ region, the noise distribution varies significantly
more rapidly than the signal distribution (see Figs. 9 and $10$ in ~\cite{Messick:2016aqy}). Therefore, the contribution of
the signal term to $\mathcal{L}$ is approximately constant over this
region, and the FAR is mostly determined by the noise distribution.

\item Finally, although $\mathcal{L}$ depends on parameters beyond $\rho$
and $\xi$, namely the event time, phase, mass, and spin, 
those should be minimally affected by the kinds of GR deviations that we consider here.
Since the polarizations are unaffected by phasing corrections and
the signals still propagate at the speed of light, the expected distribution of time delays and
phase differences across detectors will remain the same. 
Regarding masses and spins, it is possible for
non-GR signals to be identified by GR templates with masses and spins that
differ from the source. Though this would change the population model's
contribution, the model itself is broad (see Section IVB of the
GWTC-2 publication~\cite{LIGOScientific:2020ibl}) and contributes weakly to the overall
value of $\mathcal{L}$.

\end{enumerate}
These three reasons justify our $\bar{\rho}$ criterion as a proxy for detecting signals with high significance.

\section{Impact on detection efficiency}
\label{sec:singleinj}

%
\begin{figure*}
\centering
\includegraphics{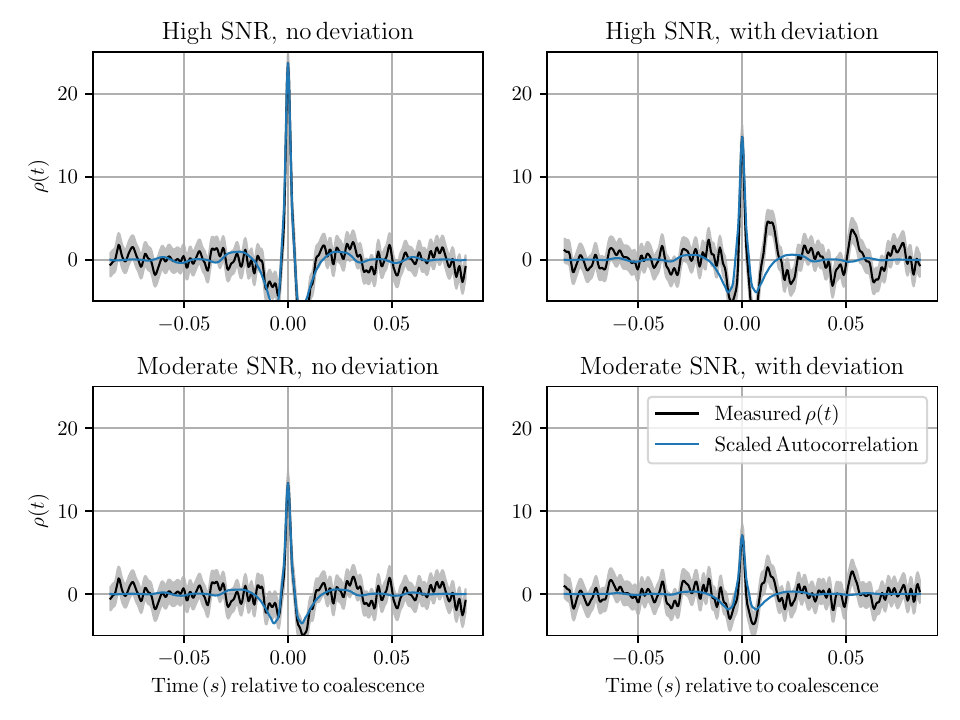}
\caption{The response of a single search template to a $30 M_\odot - 30 M_\odot$ BBH without (left) and
with (right) deviations to $\delta \varphi_{-2}$ for SNR $\sim 24$ (top) and
$\sim 15$ (bottom) injections in Gaussian noise colored to O3 sensitivities. The injections that deviate from GR use $\delta \varphi_{-2} = -0.1$.
The black line shows the measured SNR time series for a single template
waveform,
with the gray band denoting the 1$\sigma$ measurement uncertainty. The
beyond-GR phasing results in an SNR loss of $\sim40\%$ between the left and
right columns. Additionally, there is a
mismatch between the measured SNR time series and the SNR scaled
autocorrelation that weakens the signal consistency test, $\xi^2$.
Both effects lead to a reduction of our detection statistic $\bar{\rho}$, Eq.~\eqref{eq:detstatistic}, and thus a loss in sensitivity.
}
\label{fig:efficiencydemo}
\end{figure*}

To develop intuition for how deviations in the PN parameters affect the detection statistic, $\bar{\rho}$ in Eq.~\eqref{eq:detstatistic}, Fig.~\ref{fig:efficiencydemo} shows the SNR and autocorrelation time series with (right) and without (left) a deviation applied to the $-1$PN coefficient, $\delta \varphi_{-2}$, for a high (top) and low (bottom) injected SNR.
We examine these two ingredients of the total detection
statistic $\bar{\rho}$ for a characteristic BBH with redshifted masses $30{-}30\, M_{\odot}$ in
the detector frame. 
The two components of $\bar{\rho}$, $\rho$ and $\xi^2$, are represented in these plots by, respectively, the peak of the SNR time series (black) and the integrated area between it and the scaled autocorrelation time series (blue).
Mismatches between a signal and the template bank induced by a GR deviation will impact detection
efficiency due to both a loss in the recovered SNR $\rho$ (reduction in the peak height) and increase in the signal
consistency check value $\xi^2$ (increased disagreement between blue and black curves).

Indeed, the beyond-GR deviation causes a reduction in the recovered SNR, seen through a
reduced peak between the left and right panels of Fig.~\ref{fig:efficiencydemo}, thus directly affecting
$\bar{\rho}$. Moreover, the introduction of beyond-GR effects creates secondary peaks in the
SNR time series obtained from filtering with a GR waveform. The oscillations in
SNR further reduce the signal consistency check, $\xi^2$---that is, the
square difference between the measured SNR and the scaled autocorrelation, per
Eq.~\eqref{eq:consistency-def}. These oscillations become harder to discern from the Gaussian background 
with decreasing SNR, thus
minimizing the effect of $\xi^2$ on the detectability of the signal.
Figure~\ref{fig:efficiencydemo} is helpful in understanding the interplay between $\rho$ and $\xi^2$ in the presence of a deviation from GR.
However, it is not sufficient to determine the degree of selection bias against beyond-GR signals, as it only shows the effect of a single injection relative to the corresponding GR template with the same parameters.
In an actual search, we compare a beyond-GR injection against the entire bank, and the detection statistic is based on the best match.

\begin{figure*}
	\centering
	\includegraphics[width=\linewidth]{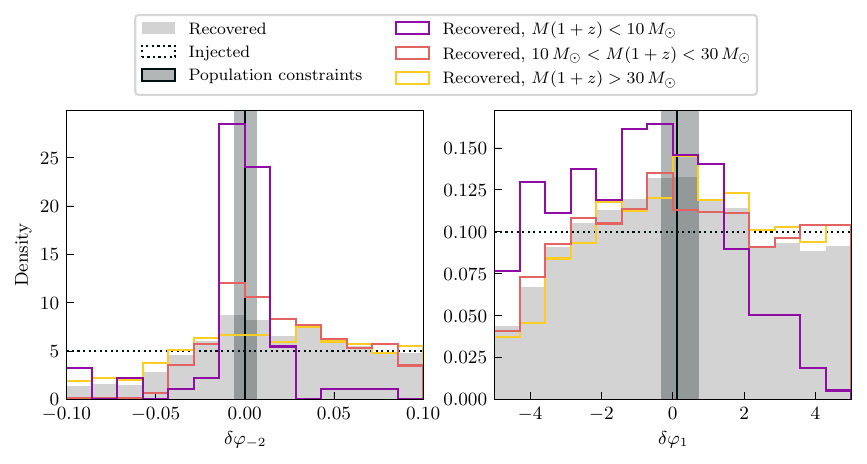}
	\caption{Histograms of recovered injections with deviations from GR in the -1PN ($\delta\varphi_{-2}$, left) and 0.5PN ($\delta\varphi_{1}$, right) coefficients.
  Although the initial injection set was assigned deviations from a uniform distribution (dotted black), the pipeline selects against large negative values of the deviation parameters, as indicated by the dearth of detections in the leftmost bins (gray histograms).
  Besides the total set of injections, we show sub-distributions corresponding to different injected mass bins in the detector frame (colored histograms). The distributions of recovered injections are largely flat over the span of values allowed by the analysis of the 12 events considered in Sec.~\ref{sec:population} (which are ${\sim}4\times$ broader than GWTC-3 constraints \cite{LIGOScientific:2021sio}; vertical gray band, median and 90\% CL), suggesting that the selection bias is not strong enough to affect the population constraints.
  }
	\label{fig:detectionhistogram}
\end{figure*}

To quantify the actual impact of GR deviations on the detection efficiency, we study the distribution of parameters of the signals that made it through our simplified detection pipeline, i.e., those that returned a value of $\bar{\rho} > 10$ when compared against \emph{any} template in the GR bank. This amounts to measuring the detectable fraction,
\begin{equation}
\hat{\mathcal{E}}(\Lambda) = \int\mathrm{d}\theta\, p_{\mathrm{det}}(\theta)\, \pi(\theta | \Lambda) \,,
\end{equation}
where $\Lambda$ is the set of hyperparameters that describe the underlying population distribution, $\pi(\theta|\Lambda)$, and $p_\mathrm{det}(\theta)$ is the selection function that describes the probability of detecting a system with parameters $\theta$.
Figure~\ref{fig:detectionhistogram} shows the marginal
selection function, $p_\mathrm{det}(\delta \varphi)$, for the -1PN coefficient
($\delta \varphi_{-2}$, left) and the 0.5PN coefficient ($\delta\varphi_1$,
right), over the whole mass space (gray) as well as subsections for different
BBH mass bins (colors).
For both parameters, the distribution of detected signals departs from the uniform intrinsic distribution that we injected (black): there is a dearth of detected signals with large negative values of the deviation parameters, indicating that such signals are selected against.
This can be explained by the fact that a negative value for these parameters will shorten the inspiral, which in turn reduces the SNR of the signal.
This effect is more pronounced for the $-1$PN coefficient, which is consistent with the intuition that this coefficient should have a larger impact on the GW phase than the 0.5PN coefficient over the duration of an inspiral because it is associated with a correction entering at a lower power of the frequency.
The drop in detection efficiency is also sharper for lower masses, as expected given the scaling of the inspiral length
with the BBH mass.

In spite of the drop in sensitivity observed at the edges of the histograms in Fig.~\ref{fig:detectionhistogram}, the recovered distributions are generally flat in the region that is allowed by the population constraints from GWTC-3 (gray band).
Lower detector-frame masses demonstrate a larger gradient across these regions (e.g. $M(1+z) < 10\,M_\odot$; purple). 
However, the observed events considered here do not reside in this region of the mass parameter-space. 
Since there is no gradient in the region allowed by the observations, there is no preference for any particular value of the deviation parameter in the range still consistent with current data. 
This suggests that the selection bias is not strong enough to affect the population constraints, which are more sensitive to GR deviations than the detection pipeline.
We confirm this below by repeating catalog analysis of GR deviations with and without the selection effects.

\section{Updated population estimates}
\label{sec:population}

We incorporate the selection function computed from Sec.~\ref{sec:singleinj} into population-level inference for inspiral tests of GR.
By computing the astrophysical distribution of beyond-GR parameters, we can now make statements about the types of GR deviations consistent with an observed set of detections.
In practice, computing the astrophysical distribution requires incorporating
knowledge of the detection efficiency over parameter space to deconvolve the
instrument's selection function from the set of observed measurements.

We evaluate the consistency of a set of observations with GR through a hierarchical analysis without imposing strong assumptions about the nature of the deviation across events.
As a null test, we follow \cite{Isi:2019asy,Isi:2022cii,Payne:2023kwj,LIGOScientific:2021sio} in parameterizing the intrinsic distribution of individual-event values for some deviation parameter $\delta\phi$ as a Gaussian $\delta\phi \sim \mathcal{N}(\mu, \sigma)$.
This model targets the mean $\mu$ and variance $\sigma^2$ of GR deviations, regardless of the true shape of the underlying distribution.
Beyond-GR parameters are typically defined to vanish in GR, so that the null hypothesis that GR is valid for all events predicts $\mu = \sigma = 0$.
If GR is not correct, then the deviation parameters may take different (nonzero) values as a function of source parameters, resulting in nonvanishing $\mu$ or $\sigma$.
We apply the approach in \cite{Payne:2023kwj} to simultaneously model the distribution of astrophysical parameters.

Existing implementations of this hierarchical analysis characterize the set of \emph{observed} events but do not inform about possible \emph{intrinsic} deviation distributions that predict events with such large deviations that are undetectable.
To factor this in, we use the result of Sec.~\ref{sec:singleinj} following the techniques used in the context of astrophysical inference to study the astrophysical distribution of within-GR parameters, such as masses and spins.
The key additional step is to incorporate the detection efficiency into the hierarchical likelihood through a term that can be approximated as the Monte-Carlo sum population weights over a set of $m$ detected injections with parameters $\theta_k$~\cite{Farr:2019rap,Miller:2020zox,Essick:2022ojx},
\begin{equation}
\label{eq:selectionMC}
\hat{\mathcal{E}}(\Lambda) = \frac{1}{M} \sum_k^m \frac{\pi(\theta_k | \Lambda)}{p(\theta_k | {\rm draw})} \,,
\end{equation}
where $M$ is the total number of drawn injections (out of which $m$ were detected), $p(\theta_k|{\rm draw})$ is the probability of drawing parameters $\theta_k$ from the population adopted in the injection campaign, with $\Lambda = \{\mu, \sigma\}$, in addition to the parameters describing the astrophysical population of GR quantities (like masses and spins).
The hierachical likelihood, $p(\{d\}|\Lambda)$, governing the inferred astrophysical population from $N$ observations with dataset $\{d\}$ is
\begin{equation}\label{eq:hierlike}
p(\{d\}|\Lambda) = \frac{1}{\hat{\mathcal{E}}(\Lambda)^N}\prod_i^N\int\mathrm{d}\theta_i\,p(d_i|\theta_i)\pi(\theta_i|\Lambda)\,,
\end{equation}
where $p(d_i|\theta_i)$ are the individual event likelihoods. 
The selection function influences the inferred hyperparameters through its inclusion in Eq.~\eqref{eq:hierlike}.

In order to include an injection in the ``detected" sum of Eq.~\eqref{eq:selectionMC}, besides \textsc{GstLAL}'s detection threshold of $\bar{\rho}>10$ from Sec.~\ref{sec:gstlal}, we additionally require that the measured SNR in the inspiral satisfy $\rho_{\rm insp} > 6$. 
The latter corresponds to the selection criterion for estimating the inspiral PN coefficients in~\cite{LIGOScientific:2019fpa, LIGOScientific:2020tif, LIGOScientific:2021sio}.
In order to avoid computing the inspiral SNR for each injection in the set, we approximate the fraction of SNR in the inspiral as a linear function of the detector frame total mass as in \cite{Payne:2023kwj}.

In addition to hierarchically modeling the beyond-GR astrophysical distribution, we incorporate population models for the within-GR population distributions. 
Due to a lower number of recovered injections than the standard set of injections used in population studies~\cite{Essick:2022ojx}, we only infer the primary mass and mass ratio distributions jointly with the beyond-GR population, using the models outlined in Ref.~\cite{Payne:2023kwj}. 
We fix the spin distribution to be uniform in spin-magnitude and isotropic about all possible spin orientations; the redshift distribution is consistent with the \textit{maximum a posteriori} power-law found in Ref.~\cite{KAGRA:2021duu}. 

With the setup described above, we repeat the hierarchical analysis in~\cite{Isi:2019asy,LIGOScientific:2020tif,LIGOScientific:2021sio,Payne:2023kwj} applied to 
12 events in O3a, to be consistent with times over which the selection function is estimated.
A list of the included events can be found in Table I of Ref.~\cite{Payne:2023kwj}. 
Figure \ref{fig:dchiMinus2-corner} shows the resulting inference on $\mu$ and $\sigma$ for the $-1$PN coefficient, $\delta\varphi_{-2}$, compared to the result that does not account for selection biases in the beyond-GR parameters.
Although this was the coefficient with the strongest detection bias as evaluated in the previous section (Fig.~\ref{fig:detectionhistogram}), this effect is very small, and the two results, with and without selection, are consistent with each other up to a slight widening of the population when selection is factored in.
This is consistent with the expectation from Fig.~\ref{fig:detectionhistogram}, which suggested the impact of selection should be minimal in light of the accuracy of the constraint from parameter estimation.
Figure \ref{fig:violins} shows that this is the case for all coefficients, none of which show significant differences between the two results.

\begin{figure}
\includegraphics[width=\columnwidth]{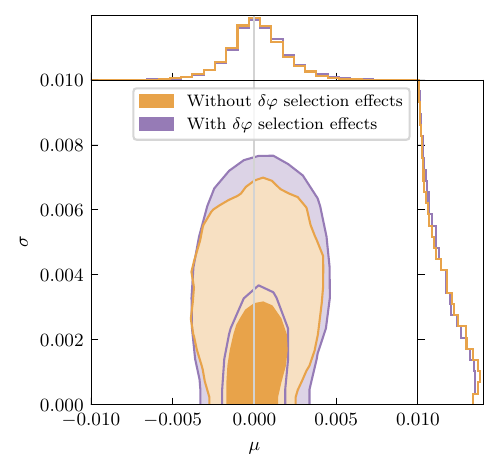}
\caption{Inference on the mean and standard deviation of the $-1$PN coefficient, $\delta\varphi_{-2}$. The orange contours show the result of the hierarchical analysis without accounting for selection effects, while the purple contours show the result when the selection function is included. The two results are consistent with each other, with the selection function widening the population only slightly.
We find no difference in the coupling between $\mu$ and $\sigma$ and the parameters controlling the mass distribution either (not shown).
}
\label{fig:dchiMinus2-corner}
\end{figure}

\begin{figure*}
\includegraphics[width=\linewidth]{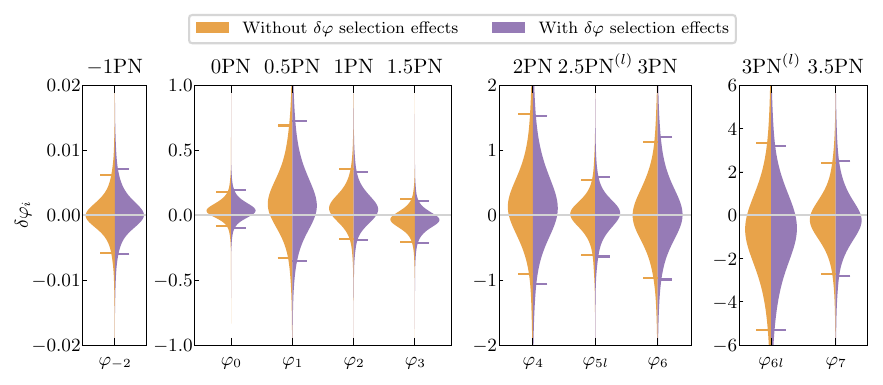}
\caption{Posterior predictive distributions (also known as the population-marginalized expectation) for deviations at all PN orders we consider, without (orange) and with (purple) selection effects factored in. No coefficient shows a significant impact when factoring in the selection: the $\delta\varphi_{-2}$ displays the strongest effect, with a slight broadening of the inferred distribution at the level of ${\sim}10\%$.}
\label{fig:violins}
\end{figure*}

\section{Conclusions}
\label{sec:conclusions}

In this study, we revisited tests of GR from the inspiral GW phase by accounting for the selection effect of templated searches against signals with GR deviations. 
We estimated the selection function by considering the performance of a simplified version of the \textsc{GstLAL} search pipeline against simulated signals with beyond-GR effects affecting the PN evolution of a BBH inspiral. 
Since \textsc{GstLAL} detects signals by comparing them to a template bank constructed with GR waveforms, 
 its detection efficiency decreases under sufficiently large deviations from GR. 
However, we found that this threshold for deviations is less stringent than the precision of GWTC-3 constraints, suggesting that 
population inference on the inspiral deviation parameters is minimally affected by selection effects.
In other words, existing constraints are already a very good approximation to the full astrophysical
population of deviation parameters, apart from the possibility of a disconnected subpopulation of sources with very high deviations.

This finding can be understood by noting that the sensitivity of parameter estimation to deviations from GR scales inversely with the SNR of the signal, while the detection threshold imposed by the search pipelines is best represented as a hard SNR cutoff.
A deviation $\delta \varphi$ that induces a mismatch $\mathcal{M}$ relative to the best-fitting GR template will result in an SNR loss of order $\rho \to \mathcal{M}\, \rho$; accordingly, the measurement precision in parameter estimation will scale as $\Delta (\delta \varphi) \sim 1/\rho$.
For a given SNR, the mismatch tolerated by the search pipeline will be much higher than the sensitivity of the parameter estimation.
Therefore, signals that incur an SNR penalty would still be detectable as long as they remain above the search's threshold; meanwhile, given a GR signal in the data, parameter estimation will constrain the magnitude of a deviation tightly around zero, with much better precision than would be directly associated with the pipeline's detection threshold.

In other words, the tolerance for detection is much larger than the tolerance for parameter estimation, and the latter is what determines the population constraints.
Since the population of observed deviations is extremely narrow (a delta function at zero if GR is correct), the hierarchical measurement is minimally affected by selection effects, as we have shown in Fig.~\ref{fig:violins}.
This argument does not apply to other parameters, such as the BH masses, since their distribution is intrinsically broad.

Our main conclusion is that the deviation population is already narrower than the extent of the selection effects,
and thus the latter do not impact the former. However, this assumes that deviations form a single, compact population
whose mean and standard deviation we constrain. Since no observed events are inconsistent with GR, 
the inferred width of this population grows smaller as the catalog increases. We are therefore not considering,
and thus not ruling out, disjoint populations with a subset of events that have extremely large (and potentially undetectable)
deviations or a mass-dependent deviation population model.
It remains conceivable that a subpopulation of signals with extremely high deviations could exist and remain hidden from GR-based pipelines, motivating dedicated searches~\cite{Chia:2020psj,Chia:2023tle,Narola:2022aob}.
However, that does not translate into selection biases for the components of the population that are already constrained by the existing catalog.

This distinction also suggests that there is no contradiction between our results and those of Ref.~\cite{Chia:2020psj,Chia:2023tle,Narola:2022aob}: we both find appreciable selection effects for sufficiently large values of the deviation parameters, c.f., Fig.~\ref{fig:detectionhistogram}. Our study, however, highlights that under the assumption of a single, unimodal population distribution of the deviation parameters, such large values of the deviation parameters are already ruled out.

As \citet{Essick:2023upv} recently pointed out, the existence of prominent selection biases would complicate the interpretation of hierarchical constraints that do not factor in selection effects, as the inferred population would not be strictly representative of neither the true astrophysical distribution nor the observed distribution of parameters.
However, in the absence of strong selection effects, hierarchical inference \emph{without} a selection term remains a valid tool to constrain the population of beyond-GR parameters, as we have shown here for PN tests of the BBH inspiral.
This, of course, may not be the case for other tests or implementations.

Our results are subject to a number of caveats, and selection effects might be stronger for different GR tests
or population models.
First, to mitigate computational costs, we have used an approximate ranking statistic
that only incorporates information from a single detector. We impose a
detection threshold of $\rho \geq \bar{\rho} \geq 10$ to maximize purity in accordance with
the FAR threshold adopted in past GR tests~\cite{LIGOScientific:2019fpa, LIGOScientific:2020tif, LIGOScientific:2021sio}.
We do not expect a full injection campaign utilizing the complete ranking
statistic described in~\cite{Tsukada:2023edh} would yield more precise results
at this threshold and for the inspiral deviation test considered here.
However, our results do not obviate the need for a full injection campaign for other tests of GR or other pipelines.

Besides the adopted threshold, the $\bar{\rho}$ ranking statistic differs from the full likelihood ratio
also on the information it considers. The latter also includes information about the phase and time of the signal in
different detectors. Though we do not expect those terms to be important for the inspiral deviation parameters
we consider here, they could become important for other tests of GR, such as those considering propagation effects
or the signal polarization. Quantifying selection effects for such tests would require a full 
multi-detector and likelihood ratio calculation.

We produce injected signals with GR deviations using standard infrastructure \cite{Husa:2015iqa,Khan:2015jqa,Hannam:2013oca,Li:2012a,Li:2012b,Agathos:2013upa}, and choose parameter ranges consistent with priors used in LIGO-Virgo-KAGRA publications.
However,
for some of these extreme values, the resulting waveform could become pathological \cite{Johnson-McDaniel:2021yge}, and may not represent a physically meaningful configuration \cite{Perkins:2022fhr}.
Although this might affect the overall applicability and physical interpretation of the tests, it does not
affect the interpretation of our results that relate to the selection effects of the tests as formulated.
Reformulations of the inspiral tests to ensure the GW phase calculation remains in the convergent series expansion
regime~\cite{Perkins:2022fhr,Wolfe:2022nkv} would likely be affected by selection effects even less, as they restrict the allowed range of
possible deviations.
 
Among the compact-binary pipelines, we restrict to a simplified version of \textsc{GstLAL}. We
expect the impact of this assumption to be small, as we only consider the most
confidently-detected BBHs with single detector SNRs $\gtrsim 10$, all of which are
detectable by \textsc{GstLAL}.
If we decreased the SNR threshold, we might
encounter events detected by other compact-binary pipelines, in which case we
would need to quantify their selection effects.
However, we expect that relaxing SNR or FAR thresholds should only make pipelines more tolerant to signals beyond GR.

Extending beyond matched-filter pipelines, we expect weakly-modeled
search methods~\cite{Klimenko:2015ypf,Cornish:2020dwh} to surpass
template-based ones for sufficiently large GR deviations. However, it is the case that both all
events we consider here and all events that have been detected in general are detected
significantly by at least one template-based search. Ultimately, the sensitivity
of weakly-modeled searches should also be quantified and taken into account, though some have
started to explore the biases this would introduce~\cite{Narola:2022aob}.

As the sensitivity of GW detectors improves, so does the number and quality of detections, leading to increasing sensitivity to both subtle deviations from GR and systematics in our models.
While here we have focused on tests of GR based on GW inspiral phases and single-Gaussian populations, exploring the effect of selection biases in other tests or under other population models will also become important. 
As both our detectors and techniques evolve, future studies need to evaluate this and other potential systematics.

\begin{acknowledgments}
We thank Reed Essick for helpful discussions.
We thank Leo Tsukada for discussions on \textsc{GstLAL}.
The Flatiron Institute is funded by the Simons Foundation.
KC was supported by NSF Grant PHY-2110111.
GP gratefully acknowledges support from a Royal Society University Research Fellowship URF{\textbackslash}R1{\textbackslash}221500 and RF{\textbackslash}ERE{\textbackslash}221015.
LIGO was constructed by the California Institute of Technology and
Massachusetts Institute of Technology with funding from the National Science
Foundation and operates under cooperative agreement PHY-1764464.
This research has made use of data, software and/or web tools obtained from the Gravitational Wave Open Science Center (https://www.gw-openscience.org), a service of LIGO Laboratory, the LIGO Scientific Collaboration and the Virgo Collaboration.
Virgo is funded by the French Centre National de Recherche Scientifique (CNRS), the Italian Istituto Nazionale della Fisica Nucleare (INFN) and the Dutch Nikhef, with contributions by Polish and Hungarian institutes.
This material is based upon work supported by NSF's LIGO Laboratory which is a major facility fully funded by the National Science Foundation.
The authors are grateful for computational resources provided by the LIGO Laboratory and supported by NSF Grants PHY-0757058 and PHY-0823459.
This research has made use of data or software obtained from the Gravitational
Wave Open Science Center (gwosc.org), a service of LIGO Laboratory, the LIGO
Scientific Collaboration, the Virgo Collaboration, and KAGRA. 
This paper carries LIGO document number \dcc{2300381}.
The filtering was performed with the \textsc{GstLAL}
library~\cite{Tsukada:2023edh,2021SoftX..1400680C,Sachdev:2019vvd,Messick:2016aqy},
built on the \textsc{LALSuite} software library~\cite{lalsuite}. Our
hierarchical analysis utilizes
NumPyro~\cite{phan2019composable,bingham2019pyro}, JAX~\cite{jax2018github},
AstroPy~\cite{astropy:2013,astropy:2018,astropy:2022},
Numpy~\cite{harris2020array}, and SciPy~\cite{2020SciPy-NMeth}. The plots shown
in this work use Matplotlib~\cite{Hunter:2007}, seaborn~\cite{Waskom2021},
arViz~\cite{arviz_2019}, and corner~\cite{corner}.
\end{acknowledgments}


\bibliography{cbc-group}

\end{document}